\documentclass[notoc]{JHEP3}
\usepackage{amsfonts,amssymb,amsmath,cite}
\title{Veneziano-Yankielowicz Superpotential Terms in $\mathcal{N}=1$
  SUSY Gauge Theories} \author{Ben M. Gripaios and John F. Wheater
  \\Department of Physics - Theoretical Physics, University of Oxford,
\\1, Keble Road, Oxford, OX1 3NP,  UK.
Fax. +44 1865 273947 
\\ e-mail: \email{b.gripaios1@physics.ox.ac.uk} ,
\email{j.wheater1@physics.ox.ac.uk}}
 \abstract{The Veneziano-Yankielowicz glueball
  superpotential for an arbitrary $\mathcal{N}=1$ SUSY pure gauge
  theory with classical gauge group is derived using an approach
  following recent work of Dijkgraaf, Vafa and others. These
  non-perturbative terms, which had hitherto been included by hand,
  are thus seen to arise naturally, and the approach is rendered
  self-contained. By minimising the glueball superpotential for
  theories with fundamental matter added, the expected vacuum
  structure with gaugino condensation and chiral symmetry breaking is
  obtained. Various possible extensions are also discussed.}
\preprint{OUTP-03 20P} 
\begin{document}
\section{Introduction}
Recent work following a conjecture by Dijkgraaf and Vafa
\cite{Dijkgraaf:2002dh} has shown that non-perturbative information
about the vacuum structure of $\mathcal{N}=1$ SUSY gauge theories with
arbitrary matter can be obtained via perturbative planar diagram
computations in a matrix model.

The gauge theory/matrix model correspondence, originally established
via a chain of dualities in string theory (in which the gauge theory
is embedded)
\cite{Bershadsky:1994cx,Cachazo:2001jy,Dijkgraaf:2002fc,
Dijkgraaf:2002vw,Dijkgraaf:2002dh}
is ultimately purely field-theoretic. A diagrammatic proof has been
supplied in \cite{Dijkgraaf:2002xd}; a proof based on demonstrating
the equivalence between Ward identities following from a generalized
Konishi \cite{Konishi:1984hf,Konishi:1985tu} anomaly of the gauge
theory (on the one side) and the loop equations of the matrix model
(on the other) was given in \cite{Cachazo:2002ry}.

To be precise, only a part of the contributions to the glueball
superpotential have been calculated (thus far) using the matrix model
correspondence; there is an additional non-perturbative contribution
coming from Veneziano-Yankielowicz terms
\cite{Veneziano:1982ah,Taylor:1983bp}. Previously, these have been
included `by hand', though in the original matrix model approach, it
was noted that these terms can come from the matrix model measure
\cite{Ooguri:2002gx,Dijkgraaf:2002dh}; in the Konishi anomaly
approach, they correspond to an undetermined constant of integration
\cite{Cachazo:2002ry}.

It is shown here that the Veneziano-Yankielowicz terms, which are
non-perturbative contributions coming from the gauge fields, can in
fact be derived in the Dijkgraaf-Vafa context (thus rendering the
approach self-contained) by the following argument. One considers the
case with a classical gauge group and flavours of matter in the
fundamental representation. The vacuum structure of such theories has
been known for some time \cite{Affleck:1984mk}, and indeed it is
re-derived below. If the matter fields have non-zero expectation
values in the vacuum, then the gauge group is spontaneously broken at
low energies via the Higgs mechanism.

The tree-level matter superpotential is such that, classically, there
are vacuum branches in which some matter fields have zero \emph{vevs},
whilst others have non-zero \emph{vevs}. This allows the gauge
symmetry breaking to be engineered. However, as shown in
\cite{Argurio:2002xv,Brandhuber:2003va}, the tree-level matter
superpotential is sufficiently simple that the effective glueball
superpotential can be determined exactly from the standard Konishi
anomaly Ward identity (up to some constant of integration independent
of the the matter couplings in the tree-level superpotential). By
considering two different vacua, with two different low energy gauge
groups, a difference equation is obtained whose solution yields the
Veneziano-Yankielowicz superpotential for the low energy pure gauge
theory.

In the next section, this is performed for the gauge group $SU(N)$.
Once the low energy effective superpotential for the pure gauge theory
(the Veneziano-Yankielowicz part) has been obtained, the full
superpotential (with the constant of integration determined) can be
obtained \cite{Brandhuber:2003va} by matching it to any one of its low
energy pure gauge theory limits (in which all the matter is integrated
out). The vacuum structure is then determined by finding the critical
points of the glueball superpotential; the expected pattern of gaugino
condensation and chiral symmetry breaking is observed. In section
\ref{sosp}, the extension to other classical gauge groups is
performed. Again, the results are as expected. In section \ref{mm},
the connection with the matrix model is outlined and in section
\ref{disc}, the results and various possible extensions are discussed.
\section{Special Unitary Groups}
\label{SUN}
Consider the $\mathcal{N}=1$ supersymmetric $SU(N)$ gauge theory in
four dimensions with glueball chiral superfield
$S=-\frac{1}{32\pi^2}\mathrm{tr} W^{\alpha}W_{\alpha}$.  With chiral
matter superfields $\Phi$ added, the theory has the Konishi
anomaly\cite{Konishi:1984hf,Konishi:1985tu,Cachazo:2002ry}. For the
chiral change of variables $\delta \Phi = \epsilon \Phi' (\Phi)$, one
has
\begin{gather} \label{konishi}
  \left \langle \Phi'_{I} \frac{\partial W_{\mathrm{tree}} }{\partial
      \Phi_I}+ \left( \frac{1}{32\pi^2} W_{\alpha J}^{K} W^{\alpha
        J}_{I} \right) \frac{\partial \Phi'_{K}}{\partial \Phi_{I}}
  \right \rangle = 0,
\end{gather}
where the indices carry the representation of the gauge group and the
tree-level matter superpotential $W_{\mathrm{tree}} = g_k \Phi^k$ is
some (gauge- and flavour-invariant) polynomial in the matter
superfields. The set of such Ward identities can be solved for the
vacuum expectation values of the matter superfields in a background
consisting of the light degrees of freedom. The matter is assumed to
be massive and so the only light degrees of freedom are the massless
gauge superfields. One can then determine the effective superpotential
for the massless gauge superfields by solving the partial differential
equations
\begin{gather} \label{pdes}
  \frac{\partial W_{\mathrm{eff}} }{\partial g_k} = \langle \Phi^k
  \rangle,
\end{gather}
which follow by holomorphy and supersymmetry.

This paper considers the case where the matter sector consists of $F$
`quark' flavours, viz.\ $F$ chiral superfields $Q^{i}_{I}$ in the
fundamental representation and $F$ chiral superfields
$\tilde{Q}_{j}^{J}$ in the anti-fundamental, where $i$ and $j$ are
flavour indices and $I$ and $J$ are colour indices. The tree-level
matter superpotential (for $F<N$) is written in terms of the $F \times
F$ gauge-invariant meson matrix $M_{j}^{i}=Q^{i}_{I}
\tilde{Q}_{j}^{I}$ as
\begin{gather} \label{wtree}
  W_{\mathrm{tree}} = m \mathrm{tr} M - \lambda \mathrm{tr} M^2.
\end{gather}
This superpotential is non-renormalizable, but this is irrelevant. It
can be obtained from a renormalizable superpotential with an
additional adjoint matter superfield by integrating out the extra
matter.  The classical equations of motion for the matter fields are
\begin{gather} \label{classeom}
  mM_{i}^{j} - 2\lambda M_{i}^{k}M_{k}^{j} = 0.
\end{gather}
The meson matrix $M$ can be brought to diagonal form via a global
flavour transformation; then the classical vacua have $F_{-}$
eigenvalues at $M_{i}^{i}=0$ and $F_{+}=F-F_{-}$ eigenvalues at
$M_{i}^{i}= m/2\lambda$ (no sum on $i$), with the low energy gauge
group broken down to $SU(N-F_{+})$.

The classical dynamics is modified by quantum effects. Consider the
transformation $\delta Q_{I}^{i} = \epsilon Q_{I}^{j}$. From
(\ref{konishi}), this yields the anomalous Ward identity
\begin{gather} \label{kontwo}
  \left\langle mM_{i}^{j} - 2 \lambda M_{i}^{k} M_{k}^{j} +
    \delta_{i}^{j} \frac{1}{32\pi^2} W_{\alpha J}^{I} W^{\alpha J}_{I}
  \right\rangle =0.
\end{gather}
This is to be evaluated in a background consisting of the massless
gauge superfields. At the classical level, it was seen that the gauge
group is broken via the Higgs mechanism, with gauge superfields
corresponding to broken generators becoming massive. The background
should only contain the superfields corresponding to the unbroken
generators.

Let the corresponding background glueball superfield be denoted
${S'}$. The third term in the vacuum expectation value in
(\ref{kontwo}) splits into two parts. The part tracing over the
unbroken gauge group is by definition ${S'}$. The other part traces
over the broken part of the gauge group. The associated superfields
are massive, and their potential is (classically) quadratic and
centred at the origin, such that their vevs are zero.\footnote{This is
  certainly true at the classical level, but it is possible that
  quantum corrections will modify this. However, the limit will be
  taken later later on in which the masses of the gauge bosons go to
  infinity and they decouple. In this limit, their vevs certainly are
  zero, and so the possibility of quantum corrections will not affect
  the argument below.} Furthermore, the matter expectation values
factorise and so (\ref{kontwo}) becomes
\begin{gather} \label{quanteom}
  m \langle M_{i}^{j} \rangle - 2 \lambda \langle M_{i}^{k}M_{k}^{j}
  \rangle = \delta_{i}^{j} {S'},
\end{gather}
which represents quantum corrections to the equation of motion
(\ref{classeom}). Up to a global flavour rotation, (\ref{quanteom})
has the solution
\begin{gather} \label{}
  \langle M_{i}^{j}\rangle = \delta_{i}^{j} \frac{m}{4\lambda}\left( 1
    \pm \sqrt{1-\frac{8\lambda {S'}}{m^2}} \right),
\end{gather}
where the solution with the plus sign corresponds to the Higgsed
vacuum in the classical limit and conversely. The eigenvalues of $M$
can be distributed as before, so that
\begin{gather} \label{}
  \langle \mathrm{tr} M \rangle = F_{-} \frac{m}{4\lambda}\left( 1 -
    \sqrt{1-\frac{8\lambda {S'}}{m^2}} \right) + F_{+}
  \frac{m}{4\lambda}\left( 1 + \sqrt{1-\frac{8\lambda {S'}}{m^2}}
  \right),
\end{gather}
and similarly for $\langle \mathrm{tr} M^2 \rangle$.  The partial
differential equations for the effective glueball superpotential with
respect to the matter sector couplings are
\begin{align} \label{}
  \frac{\partial W_{\mathrm{eff}}}{\partial m} 
&= \langle \mathrm{tr} M \rangle, \nonumber \\
  \frac{\partial W_{\mathrm{eff}}}{\partial \lambda} &= - \langle
  \mathrm{tr} M^2 \rangle,
\end{align}
or
\begin{align} \label{}
  \frac{\partial W_{\mathrm{eff}}}{\partial m} &= F_{-} 
 \frac{m}{4\lambda} \left( 1 - \sqrt{1-\frac{8\lambda {S'}}{m^2}} \right)
 + F_{+}  \frac{m}{4\lambda} \left( 1 + \sqrt{1-\frac{8\lambda {S'}}{m^2}}
 \right), \nonumber \\
  \frac{\partial W_{\mathrm{eff}}}{\partial \lambda} &= - F_{-}
  \frac{m^2}{16\lambda^2}\left( 1 - \sqrt{1-\frac{8 \lambda
        {S'}}{m^2}} \right)^{2} - F_{+} \frac{m^2}{16\lambda^2}\left(
    1 + \sqrt{1-\frac{8 \lambda {S'}}{m^2}} \right)^{2} ,
\end{align}
which have the solution
\begin{multline} \label{weff}
  W_{\mathrm{eff}} = F\frac{m^2}{8 \lambda} +
  (F_{+}-F_{-})\frac{m^2}{8 \lambda}\sqrt{1-\frac{8\lambda {S'}}{m^2}}
  + F {S'} \log m \\
  + {S'} \log \left[ \left( \frac{1}{2} +
      \frac{1}{2}\sqrt{1-\frac{8\lambda {S'}}{m^2}} \right)^{F_{-}}
    \left( \frac{1}{2} - \frac{1}{2}\sqrt{1-\frac{8\lambda {S'}}{m^2}}
    \right)^{F_{+}} \right] + c({S'}).
\end{multline}
Here, $c$ is a constant of integration which must be independent of
the matter sector couplings, but may depend on other parameters, such
as ${S'}$. In order to fully specify $W_{\mathrm{eff}}$ it is
necessary to determine $c$. In \cite{Brandhuber:2003va}, this has been
done by mapping on to the known Veneziano-Yankielowicz
\cite{Veneziano:1982ah,Taylor:1983bp} form of the effective action for
the low energy gauge degrees of freedom; thus the
Veneziano-Yankielowicz terms are introduced by hand. In what follows,
it is shown that the Veneziano-Yankielowicz effective superpotential
can in fact be derived from (\ref{weff}).

To do this, first take the limit in which both the quark mass $m$ and
the gauge boson mass $\sqrt{m/2\lambda}$ become large. The effective
potential (\ref{weff}) becomes
\begin{gather} \label{w2}
  W_{\mathrm{eff}} = F_+ \frac{m^2}{4\lambda} - F_+ {S'} + F_+ {S'}
  \log \frac{{S'}}{m^2/2\lambda} + F \frac{{S'}}{2} + F {S'} \log m +
  c({S'}).
\end{gather}
What is the meaning of this expression? In this limit, the massive
degrees of freedom decouple; the effective superpotential should
consist of the superpotential for the massless gauge degrees of
freedom plus terms representing the contribution of the decoupled
matter which has been integrated out. This decoupled matter consists
of the quarks and the massive gauge bosons corresponding to the broken
generators of $SU(N)$ (and their superpartners). One can calculate the
contribution of the quarks to the effective superpotential as follows.
The non-renormalization theorem applies to the decoupled matter
sector, and the contribution is found by replacing the quark fields in
the tree-level superpotential (\ref{wtree}) by their vacuum
expectation values. The contribution to $W_{\mathrm{eff}}$ is thus
\begin{gather}
  F_{+} \frac{m^2}{4\lambda},
\end{gather}
reproducing the first term in (\ref{w2}). The contribution of the
massive gauge superfields to vevs (and therefore to the effective
superpotential) was earlier seen to be zero.  Discarding the term
independent of ${S'}$ (the contribution of the quark superfields),
what is left must represent the contribution of the massless gauge
fields alone, that is, a pure gauge theory contribution. This contains
the as yet unknown constant $c$, which can be removed by considering
the superpotentials for two distinct vacua in which the number of
Higgsed quarks, $F_+$, takes the values $F_1$ and $F_2$, but the
argument ${S'}$ takes the same value, $T$ say, in both.\footnote{Of
  course the physical interpretation of $T$ is different in the two
  vacua. Here however one simply wants to determine the functional
  form of $W_{\mathrm{eff}}$. $W_{\mathrm{eff}}$ is an unconstrained
  function of its arguments and so the arguments may be chosen
  arbitrarily.} If one then subtracts the two effective superpotential
functions, the unknown constant $c$ cancels, giving
\begin{gather} \label{dwgauge}
  \Delta W_{\mathrm{eff}} = - (F_1 - F_2) T + (F_1 - F_2) T \log
  \frac{T}{m^2/2\lambda}.
\end{gather}
This expression still involves the matter sector couplings $m$ and
$\lambda$. These account for the required matching of the scales of
the low energy $SU(N-F_{1,2})$ gauge theories (with $F_{1,2}$ Higgsed
quarks and $F-F_{1,2}$ massive quarks integrated out) to the UV scale
of the original $SU(N)$ gauge theory with $F$ flavours. Indeed, one
has \cite{Terning:2003th}
\begin{gather} \label{sunscale}
  \Lambda_{N-F_{1},0}^{3(N-F_{1})} \left( \frac{m^2}{2\lambda}
  \right)^{F_{1}} = \Lambda_{N,F}^{3N - F} m^F = \Lambda_{N-F_{2},0}
  ^{3(N-F_{2})} \left( \frac{m^2}{2\lambda} \right)^{F_{2}},
\end{gather}
where $\Lambda_{N,F}$ denotes the scale for the gauge group $SU(N)$
with $F$ flavours.  Using this relation, one can eliminate the matter
sector couplings $m$ and $\lambda$ altogether from (\ref{dwgauge}) to
obtain
\begin{multline} \label{}
  W_{\mathrm{eff}}(N-F_{1},T,\Lambda_{N-F_{1},0}) -
  W_{\mathrm{eff}}(N-F_{2},T,\Lambda_{N-F_{2},0}) = \\ (N-F_{1})
  \left( - T\log \frac{T}{\Lambda_{N-F_{1},0}^{3}} +T \right) -
  (N-F_{2}) \left( - T\log \frac{T}{\Lambda_{N-F_{2},0}^{3}} +T
  \right),
\end{multline}
where the functional dependence of $W_{\mathrm{eff}}$ has been
indicated explicitly. This difference equation has the solution
\begin{gather} \label{}
  W_{\mathrm{eff}}(N,S,\Lambda_{N,0}) = N \left( - S\log
    \frac{S}{\Lambda_{N,0}^{3}} +S \right) + f(S),
\end{gather}
where the full glueball superfield has been re-instated and $f(S)$ is
an arbitrary function of $S$ alone: it cannot depend on any of the
other parameters present. Furthermore, on dimensional grounds, $f$
must be proportional to $S$. Thus
\begin{gather}
  W_{\mathrm{eff}}(N,S,\Lambda_{N,0}) = \left( - S\log \frac{S^N}{a
      \Lambda_{N,0}^{3N}} +N S \right),
\end{gather}
where $a$ is a pure number. The arbitrariness observed in
$W_{\mathrm{eff}}$, parameterised by $a$, corresponds precisely to the
renormalisation group scheme dependence, in which one is free to shift
$\Lambda_{N,0}^{3N} \rightarrow a \Lambda_{N,0}^{3N}$.  In a scheme in
which $f$ vanishes or $a=1$, the glueball superpotential for the pure
$SU(N)$ gauge theory is
\begin{gather} \label{sunw}
  W_{\mathrm{eff}}(N,S,\Lambda_{N,0}) = N \left( - S\log
    \frac{S}{\Lambda_{N,0}^{3}} +S \right),
\end{gather}
which has precisely the form suggested by Veneziano and Yankielowicz
on the basis of extended $U(1)_R$ symmetry considerations.

Now that the effective superpotential for the low energy gauge theory
has been derived, one can determine $c$ as in \cite{Brandhuber:2003va}
by demanding that $W_{\mathrm{eff}}$ in (\ref{weff}) reproduces the
correct limit as $m^2/\lambda \rightarrow \infty$ for the vacuum with
$F_{+}$ Higgsed quarks and low energy gauge group
$SU(N-F_{+})$.\footnote{The arguments above show that if the correct
  low energy limit is obtained for one value of $F_+$, then the
  correct low energy limit will also be obtained for all values of
  $F_+$.} It is not difficult to show that the correct form is
\begin{multline} \label{weffc}
  W_{\mathrm{eff}} = {S'} \left( - \log
    \frac{{S'}^{N}}{\Lambda_{N,F}^{3N-F}m^{F}} + N \right) -
  F\frac{{S'}}{2} +F\frac{m^2}{8 \lambda}
  + (F_{+}-F_{-})\frac{m^2}{8 \lambda}\sqrt{1-\frac{8\lambda {S'}}{m^2}} \\
  + {S'} \log \left[ \left( \frac{1}{2} +
      \frac{1}{2}\sqrt{1-\frac{8\lambda {S'}}{m^2}} \right)^{F_{-}}
    \left( \frac{1}{2} - \frac{1}{2}\sqrt{1-\frac{8\lambda {S'}}{m^2}}
    \right)^{F_{+}} \right].
\end{multline}

Finally, one can show that the expected quantum vacuum structure is
reproduced.  Minimising $W_{\mathrm{eff}}$ with respect to ${S'}$, one
finds that
\begin{gather} \label{vac}
  \log \left[\frac{\Lambda_{N,F}^{3N-F}m^{F}}{{S'}^{N}}\left(
      \frac{1}{2} + \frac{1}{2}\sqrt{1-\frac{8\lambda {S'}}{m^2}}
    \right)^{F_{-}} \left( \frac{1}{2} -
      \frac{1}{2}\sqrt{1-\frac{8\lambda {S'}}{m^2}} \right)^{F_{+}}
  \right] = 0.
\end{gather}
The solution is in general non-trivial. In the limit in which quark
masses and Higgs \emph{vevs} become large however, it reduces to
\begin{gather} \label{}
  \log \left[\frac{\Lambda_{N,F}^{3N-F}m^{F}}{{S'}^{N}}\left( \frac{2
        \lambda {S'}}{m^2} \right)^{F_{+}} \right]=0,
\end{gather}
implying
\begin{gather} \label{}
  {S'}^{N-F_{+}} = \Lambda_{N-F_{+},0}^{3(N-F_{+})}.
\end{gather}
There are precisely $N-F_{+}$ vacua with gluino condensation and
chiral symmetry breaking \cite{Affleck:1984mk}.
\section{Orthogonal and Symplectic Groups}
\label{sosp}
The extension to the other classical Lie groups is straightforward.
The only differences are that \emph{i.}\ the fundamental
representation of $SO(N)$($Sp(2N)$) is (pseudo-)real, and \emph{ii.}\ 
the one-loop beta-function coefficients (and thus the scale matching
relations (\ref{sunscale})) are modified.

Because the representations are (pseudo-)real, the meson flavour
matrix can be written as $M_{ij}=Q_i {Q}_j$, where the colour indices
are implicitly contracted using the appropriate invariant tensor. The
Konishi anomaly equation is thus modified to
\begin{gather} \label{mqeom}
  2m \langle M_{ij} \rangle - 4 \lambda \langle M_{ik} M_{kj} \rangle
  = \delta_{ij} {S'},
\end{gather}
and so all equations written in section \ref{SUN} up to and including
eq. \ref{dwgauge} remain valid upon making the replacement ${S'}
\rightarrow {S'}/2$.

For the $SO(N)$ gauge theory with $F<N-4$ quarks in the fundamental
(vector) representation \cite{Intriligator:1995id}, the one-loop
coefficient of the beta-function is $3(N-2)-F$ \cite{Terning:2003th}.
The scale matching relation (\ref{sunscale}) is modified to
\begin{gather} \label{sonscale}
  \Lambda_{N-F_{1},0}^{3(N-F_{1}-2)} \left( \frac{m^2}{2\lambda}
  \right)^{F_{1}} = \Lambda_{N,F}^{3(N-2) - F} m^F =
  \Lambda_{N-F_{2},0} ^{3(N-F_{2}-2)} \left( \frac{m^2}{2\lambda}
  \right)^{F_{2}}
\end{gather}
and the low energy glueball superpotential is
\begin{gather} \label{sonw}
  W_{\mathrm{eff}}(N,S,\Lambda_{N,0}) =\left(\frac{ N-2}{2}\right)
  \left( - S \log \frac{S}{2\Lambda_{N,0}^{3}} + S \right).
\end{gather}
The only difference from the standard form is the factor of two in the
logarithm. This is however renormalisation scheme dependent.

For the $Sp(2N)$ gauge theory with $F<N+1$ flavours ($2F$ quarks)
\cite{Intriligator:1995ne}, the one-loop coefficient of the
beta-function is $3(2N+2)-2F$, whence
\begin{gather} \label{spnw}
  W_{\mathrm{eff}}(N,S,\Lambda_{N,0}) = (N+1) \left( - S \log
    \frac{S}{2\Lambda_{N,0}^{3}} + S \right).
\end{gather}
Again there is an additional factor of two in the logarithm, which is
renormalisation scheme dependent.
\section{Connection with the Matrix Model}
\label{mm}
In this section, the connection is made with the matrix model. The
first point to note is that, in the case with $F$ fundamental
flavours, the matrices $Q$ and (for $SU(N)$) $\tilde{Q}$ are $F \times
\tilde{N}$, where $\tilde{N}$ is taken to be large. Formally, the
partition function for the matrix model is
\begin{gather} \label{}
  Z = \int dQ d\tilde{Q} \exp{-\frac{1}{g_m}
    W_{\mathrm{tree}}(Q,\tilde{Q})}
\end{gather}
and the required Ward Identity (\ref{quanteom}) can be obtained
directly from
\begin{gather} \label{}
  \int dQ d\tilde{Q} \frac{d}{dQ_i} \left[ Q_j \exp{-\frac{1}{g_m}
      W_{\mathrm{tree}}(Q,\tilde{Q})} \right] = 0,
\end{gather}
upon making the replacement $g_m \rightarrow {S'}$ and noting the
factorisation of correlation functions in the large $\tilde{N}$ limit.
\section{Discussion}
\label{disc}
In the above, the Veneziano-Yankielowicz superpotential terms for pure
$\mathcal{N}=1$ gauge theories with classical gauge group have been
derived, and all results are in accord with others obtained
previously.

One can now look at possible extensions of the work presented here. A
first remark is that the pathological cases where the number of
flavours is close to the number of colours
\cite{Affleck:1984mk,Intriligator:1995id,Intriligator:1995ne,Seiberg:1994bz}
were deliberately excluded. This was initially sufficient, since one
sought only to derive the superpotentials for the pure gauge theory -
the matter sector served only to engineer the symmetry breaking.
However, one went on to determine the superpotentials and vacuum
structure of the full theory (with matter), and it would be desirable
to extend this analysis to the pathological cases. Presumably this can
be done, and would necessitate adding baryonic terms to the tree-level
matter superpotential and so on. See
\cite{Corley:2003me,Bena:2002ua,Bena:2003vk} for work already
attempted along these lines.

Secondly, it would be desirable to extend the argument to the
exceptional Lie groups. At first it would seem that an analogous
argument may work: starting with an exceptional gauge group with
fundamental matter, one can engineer a situation in which the gauge
symmetry is broken to a classical gauge group (or product thereof),
for which the Veneziano-Yankielowicz superpotential terms are known.
One could then match the full superpotential onto the known
Veneziano-Yankielowicz terms in the appropriate low energy limit.
However, there is a problem, in that the pure gauge theory
superpotentials derived herein, viz.
(\ref{sunw},\ref{sonw},\ref{spnw}) do not hold for the lowest-lying
classical Lie groups. The reason for this is clear from
\cite{Aganagic:2003xq}: the gauge group one considers in the
Dijkgraaf-Vafa framework is really the supergroup $\lim_{k \rightarrow
  \infty} SU (N+k|k)$, for which the superpotentials derived in the
present work are correct for all $N$. However, instanton effects mean
that the Veneziano-Yankielowicz terms for $SU(N)$ and its supergroup
extension are different for low-lying $N$ (consider for example
$SU(1)$). Thus, in order to derive the Veneziano-Yankielowicz terms
for the exceptional groups, one would need to fix the results for the
low-lying classical gauge groups by hand first.

Thirdly, and more speculatively, it is noted that the simple case with
fundamental matter discussed above may shed some light on the
remarkable observation that the Veneziano-Yankielowicz terms can be
obtained from the measure in the corresponding matrix model for
adjoint matter. At first thought, this seems nonsensical. The
Veneziano-Yankielowicz terms pertain to the pure gauge theory, so how
can it be that they come from the matter sector?

Take instead the viewpoint that the Veneziano-Yankielowicz terms come
from non-perturbative contributions in the pure gauge theory. Now add
matter in the adjoint representation. Just as in the case with
fundamental matter considered above, it is then possible to break the
gauge symmetry to some smaller gauge group at low energy via the Higgs
mechanism. Indeed, one can even go so far as to break the non-Abelian
part of the gauge symmetry completely using the matter sector. But if
one does this, the would-be Veneziano-Yankielowicz terms in the low
energy glueball superpotential must somehow be removed. In order to
achieve this, the matter sector must contain terms which cancel the
Veneziano-Yankielowicz terms coming from the pure gauge sector. If the
correspondence between the matter sector of the gauge theory and the
matrix model is indeed complete, then these terms ought to come from
the non-perturbative part of the matrix model, viz. the measure
factor.

All of this is pure conjecture however. In order to show it, one would
like to see how the argument presented here generalizes to matter in
other representations (in particular the adjoint) and other
superpotentials. In these more general cases, one must solve the
generalized Konishi anomaly equations in closed form. Moreover, in the
adjoint case, the spontaneous symmetry breaking cannot be engineered
in the same way: the pattern of symmetry breaking is fixed once the
form of the tree-level matter superpotential is chosen.
\acknowledgments We thank John March-Russell and the late Ian Kogan
for discussions. BMG is supported by a PPARC studentship. The related
works \cite{Ambjorn:2003rp,Merlatti:2003iy,Nakayama:2003ri} appeared
during the final stages of preparation of the manuscript.
\providecommand{\href}[2]{#2}\begingroup\raggedright\endgroup
\end{document}